\documentclass[twocolumn,amsmath,showpacs,amsfonts,aps,prc,floatfix]{revtex4}
\usepackage{graphicx}
\usepackage{bm}

\begin{document}


\title{Saturation of elliptic flow and shear viscosity}

\author{A. K. Chaudhuri}
\email[E-mail:]{akc@veccal.ernet.in}
\affiliation{Variable Energy Cyclotron Centre, 1/AF, Bidhan Nagar, 
Kolkata 700~064, India}

\begin{abstract}

Effect of shear viscosity on 
elliptic flow is studied in causal dissipative hydrodynamics in 2+1 dimensions. Elliptic flow is reduced in viscous dynamics.
Causal evolution of minimally viscous fluid ($\eta/s$=0.08), can explain the PHENIX data on elliptic flow in 16-23\% Au+Au collisions up to $p_T\approx$3.6 GeV. In contrast, ideal hydrodynamics, can explain the same data only up to $p_T\approx$1.5 GeV.  
$p_T$ spectra of identified particles are also better explained in minimally viscous fluid than
in ideal dynamics. However, saturation of elliptic flow at large $p_T$ is not reproduced.
\end{abstract}


\date{\today}  

\maketitle

%
One of the important results in Au+Au collisions is the 
large elliptic flow is non-central collisions \cite{BRAHMSwhitepaper,PHOBOSwhitepaper,PHENIXwhitepaper,STARwhitepaper}. Large elliptic flows establish that in non-central Au+Au collisions, a  collective QCD matter is created. Whether the matter can be characterized as the lattice QCD \cite{lattice} predicted Quark-Gluon-Plasma (QGP) or not,   is still a question of debate. 
Qualitatively, elliptic flow are naturally explained in hydrodynamics,
rescattering of secondaries generates pressure and drives the subsequent collective motion. In non-central collisions, the reaction zone is asymmetric (almond shaped), pressure gradient
is large in one direction and small in the other. The asymmetric pressure gradient generates the elliptic flow. As the fluid evolve and
expands, asymmetry in the reaction zone decreases and 
comes a stage when reaction zone become symmetric and system no longer generate elliptic flow.  
Elliptic flow is early time phenomena and a  sensitive probe to the early stage of the fluid.  

Ideal hydrodynamics has been quite successful in explaining RHIC data on elliptic flow \cite{QGP3}. Assuming that the collision produces QGP, which undergoes 1st order phase transition at a critical temperature $T_c$=164 MeV, elliptic flow of identified particles are explained up to $p_T\sim$1.5 GeV. Ideal hydrodynamics also reproduces the transverse  momentum dependence of identified particles, again up to $p_T\sim$ 1.5 GeV, which exhaust more than 99\% of particle production.
Success of {\em ideal} hydrodynamics 
in explaining bulk of the data \cite{QGP3}, together with the string theory motivated lower limit of shear viscosity $\eta/s \geq 1/4\pi$ \cite{Policastro:2001yc,Policastro:2002se} has led to a paradigm    \cite{Shuryak:2003xe,Heinz:2005zg} that in Au+Au collisions, a nearly perfect fluid is created.  

However, the paradigm of {\em perfect fluid}, produced in Au+Au collisions at RHIC, need to be clarified. As indicated above, the ideal hydrodynamics is only partially successful and in a limited $p_T$ range ($p_T \leq$1.5 GeV) \cite{Heinz:2004ar}. The transverse momentum spectra 
of identified particles starts to deviate form ideal fluid dynamics prediction beyond $p_T\sim$ 1.5 GeV. Experimentally determined  HBT radii are not reproduced in the ideal fluid dynamic models, the famous "HBT puzzle" \cite{Heinz:2002un}.  
It also does not reproduce the experimental trend that elliptic flow saturates at large transverse momentum. Contrary to experiment,
  ideal hydrodynamics predicts continually increasing elliptic flow.  
These shortcomings of ideal fluid dynamics indicate
greater importance of non-equilibrium (dissipative) effects  in the $p_T$ ranges greater than 1.5 GeV. Indeed, 
saturation of elliptic flow at large $p_T$ is a manifestation of non-equilibrium effects. In parton cascade simulations of relativistic Boltzmann equation, where dissipative effects are naturally included, elliptic flow 
saturates at large $p_T$ \cite{Molnar:2001ux}. However, parton cascade simulations do not reproduce experimental data unless the elastic parton cross-section is very high, $\sigma\sim$ 45 mb.  Realistic cross-sections do not generate enough flow. Recently, elliptic flow is
studied in a hybrid "hydro+cascade" model \cite{Hirano:2005xf}. In the model, initially produced QGP evolve following the ideal hydrodynamics.
Just below the phase transition, a hadronic transport model is employed for the late stage evolution. The model seems to require additional dissipation at the early quark-gluon plasma phase.
 
Dissipation in the early QGP phase can be conveniently studied in
relativistic dissipative hydrodynamics.
Theories of relativistic  dissipative hydrodynamics are well developed \cite{Eckart,LL63,IS79}. Problem of causality violation in 1st order theories \cite{Eckart,LL63}, are removed in the Israel-Stewart's  \cite{IS79} causal, 2nd order theories. Recently, there has been significant progress in numerical implementation relativistic dissipative hydrodynamics
\cite{Teaney:2003kp,Teaney:2004qa,MR04,Koide:2007kw,Baier:2006gy,Chaudhuri:2005ea,Heinz:2005bw,asis,Chaudhuri:2007zm,Romatschke:2007mq,Song:2007fn}.  Elliptic flow in causal hydrodynamics 
has been calculated in \cite{Romatschke:2007mq,Song:2007fn}.
Romatschke and Romatschke \cite{Romatschke:2007mq} concluded that while data on integrated $v_2$ is consistent with a ratio of viscosity over entropy density up to $\eta/s\approx$0.16, the data on minimum bias $v_2$ favor a much smaller 
viscosity, $\eta/s \approx$0.03, lower than the ADS/CFT bound $\eta/s$=0.08. 
A different result was obtained in \cite{Song:2007fn}. They did not compare with experiments but
minimal viscosity $\eta/s$=0.08 leads to large reduction in elliptic flow compared to ideal hydrodynamics. 
At the Cyclotron Centre, Kolkata, we have developed a code, "AZHYDRO-KOLKATA" to solve "`causal"' dissipative hydrodynamics.  In an earlier publication  \cite{Chaudhuri:2007zm}, we have discussed
the space-time evolution of QGP fluid without any phase transition.  AZHYDRO-KOLKATA results for elliptic flow (including phase transition) are presented here.

Elliptic flow of a particle (say $\pi^-$) is quantified as the 2nd harmonic of azimuthal distribution of $\pi^-$,

\begin{equation} \label{eq1}
v_2(p_T)=\frac{\int d\phi \frac{d^2N}{dyd^2p_T} cos(2\phi)}
                {\int d\phi \frac{d^2N}{dyd^2p_T}  },
\end{equation}

\noindent where $\frac{d^2N}{dyd^2p_T}$ is the invariant distribution of $\pi^-$.  In Cooper-Frye prescription,
the invariant distribution is obtained as,  
 
\begin{equation} \label{eq2}
\frac{dN}{dyd^2p_T} =\int_\Sigma d\Sigma_\mu p^\mu f(x,p),
\end{equation}

\noindent where $\Sigma_\mu$ is the freeze-out hyper surface and
$f(x,p)$ is the distribution function. 
In viscous dynamics, the system is not in (local) equilibrium. If the system is not far from equilibrium, $f(x,p)$ can be approximated as,

\begin{equation}\label{eq4}
f(x,p)=f^{(0)}(x,p) [1+\phi(x,p)],
\end{equation}

\noindent where $f^0(x,p)$ is the equilibrium distribution function,
and $\phi(x,p) << 1$, is a small correction due to non-equilibrium effects. 
With shear viscosity as the only dissipative force the correction factor
$\phi(x,p)$ can be locally approximated as, 

\begin{equation}\label{eq5}
\phi(x,p)=\varepsilon_{\mu\nu} p^\mu p^\nu =\frac{1}{2T^2(\varepsilon+p)} \pi_{\mu\nu}p^\mu p^\nu.
\end{equation}
 
\noindent where $\varepsilon$,$p$ and $T$ are the (local) energy density, pressure and temperature. $\pi^{\mu\nu}$'s  are the shear
stress tensors.

Accordingly, in viscous dynamics, $\pi^-$ invariant distribution has two parts,  

\begin{eqnarray} \label{eq6}
\frac{dN}{dyd^2p_T}
=&&\int d\Sigma_\mu p^\mu f^0(x,p)+\int d\Sigma_\mu p^\mu [f^0(x,p)\phi(x,p)] \nonumber\\
=&&\frac{dN^{eq}}{dyd^2p_T}+\frac{dN^{neq}}{dyd^2p_T},
\end{eqnarray}

\noindent  where $\frac{dN^{neq}}{dyd^2p_T}$ is the non-equilibrium correction to the $\pi^-$ 
equilibrium distribution $\frac{dN^{eq}}{dyd^2p_T}$ .
Since $\phi(x,p) << 1$ it is then necessary that $\frac{dN^{neq}}{dyd^2p_T} << \frac{dN^{eq}}{dyd^2p_T}$. 
Expanding Eq.\ref{eq1} to the 1st order, 

\begin{eqnarray}\label{eq7}
v_2(p_T) =&& v_2^{eq}(p_T) \nonumber \\&&+ \left(-v_2^{eq}\frac{\int d\phi \frac{d^2N^{neq}}{p_T dp_Td\phi}}{\int d\phi \frac{d^2N^{eq}}{p_Tdp_Td\phi}}
+\frac{\int d\phi cos(2\phi) \frac{d^2N^{neq}}{p_T dp_T d\phi}}
{\int d\phi \frac{d^2N^{eq}}{p_T dp_T d\phi}} \right) \nonumber \\
 = && v_2^{eq}(p_T) + v_2^{corr}(p_T),
\end{eqnarray} 
 
\noindent where  $v_2^{eq}$ is the  equilibrium elliptic flow  and $v_2^{corr}$ is the correction due to the non-equilibrium effects. 
Non-equilibrium correction grow quadratically with $p_T$ (see Eq.\ref{eq5}). $v_2^{eq}$ on the other hand grow less than linearly with $p_T$.   An impportant conclusion can be reached from Eq.\ref{eq7}: asympotically, viscous hydrodynamics do not predict saturation of elliptic flow.

\begin{figure}[h]
\includegraphics[bb=40 278 507 755
 ,width=0.7\linewidth,clip]{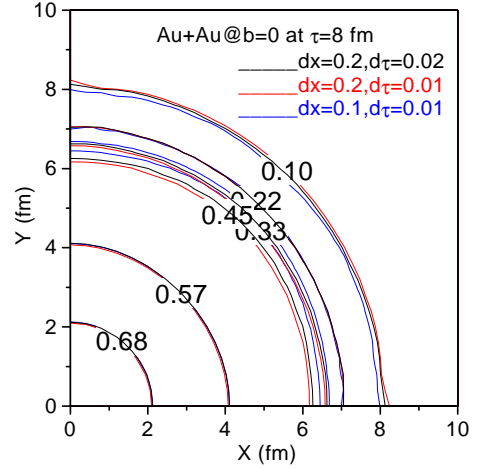}
\caption{  (color online) constant energy density contours in b=0 Au+Au collisions after 8 fm of evolution.}
\label{F1}
\end{figure}

Assuming longitudinal boost-invariance, we have solved causal dissipative hydrodynamics in 2+1 dimension. We have included the shear viscosity only. Details of equation solved and some results for space-time evolution could be found in \cite{asis}. 
To simulate the Au+Au collisions at RHIC, we have used the
standard initial conditions described in \cite{QGP3}. 
At initial time $\tau_i$=0.6 fm, the QGP fluid was initialized with central entropy density $S_{ini}=110 fm^{-3}$. It corresponds to peak energy density $\sim 35 GeV/fm^3$.
We also use 
the equation of state EOS-Q described in \cite{QGP3}, incorporating a first order phase transition at a critical temperature $T_c$=164 MeV. In viscous hydrodynamics, 
additionally, one has to initialize the shear stress-tensors. 
We assume that at initial time $\tau_i$, shear stress-tensors have attained their boost-invariant values \cite{asis}.
Viscous hydrodynamics introduces some new parameters, 
the viscosity coefficient $\eta$ and  the relaxation time $\tau_{relx}$.   Viscosity coefficient
of a QGP or hadronic fluid is quite uncertain. In a strongly coupled field theory, ADS/CFT estimate gives a lower limit to
shear viscosity, $\frac{\eta}{s} \geq \frac{1}{4\pi} \approx 0.08$ \cite{Policastro:2001yc,Policastro:2002se}. Presently we show results for the minimal shear viscosity,
$\eta/s$=0.08. For the relaxation time, we use the kinetic theory approximation for a Boltzmann gas, $\tau_{relx}=\frac{6\eta}{4p}\approx \frac{6\eta}{sT}$. 

It is very important to establish the accuracy of the numerical code. Being a ratio, elliptic flow is a sensitive observable and
any small error in computation can lead to large uncertainty in
the elliptic flow. {\em Viscous hydrodynamics do not posses analytical solution against which the numerical results could be tested}. We use the general procedures to test a numerical code, (i) the results should be stable against change in integration step lengths,  (ii) any symmetry in the system should not be destroyed and (iii) unphysical maxima or minima should not creep into due to numerical error. In Fig.\ref{F1},  we have shown the constant energy density contours in a zero impact parameter Au+Au collisions after 8 fm of evolution. The contours are drawn for three different sets of integration step lengths, (i)$dx=dy=0.2 fm,d\tau=0.02 fm$, (the black lines) , (ii)$dx=dy=0.2 fm,d\tau=0.01 fm$, (the red lines) and (iii)$dx=dy=0.1 fm,d\tau=0.01 fm$, (the blue lines). The ratio of viscosity over entropy density is $\eta/s$=0.08.
Fig.\ref{F1} shows that all the three conditions mentioned above are satisfied. The evolution is stable
against change in integration step lengths. In the interior of the fluid, energy density surfaces are exactly reproduced and one can not distinguish the lines for different integration step lengths. At large radius, energy density distribution does show dependence on integration step lengths but the dependence is small. For example, the spatial uncertainty in the  $\varepsilon$=0.1 GeV surface is less than 3-4\% only.  In a b=0 collision, initial energy density is azimuthally symmetric. The symmetry is 
maintained during the evolution. Also the smooth fall of energy density along the radius indicates that no new maxima or minima are generated during the evolution.

\begin{figure}[ht]
\includegraphics[bb=50 317 532 793
 ,width=0.7\linewidth,clip]{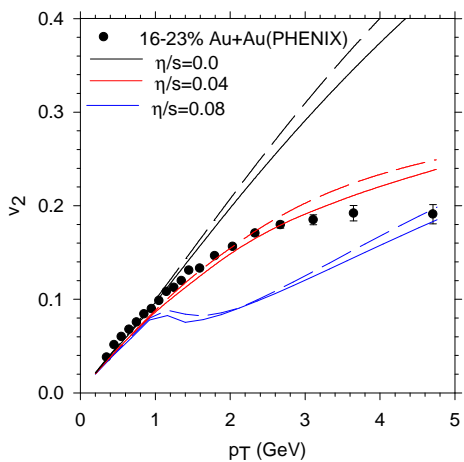}
\caption{  (color online) The solid and dashed lines show the $p_T$ dependence  of elliptic flow in evolution with integration step lengths $dx=dy=0.2,d\tau=0.02$ and $dx=dy=0.1,d\tau=0.01$.  The black lines corresponds to ideal fluid. The  red and blue lines corresponds to viscous fluid with $\eta/s$=0.04 and with $\eta/s$=0.08 respectively.  
The filled circles are PHENIX data \cite{Adler:2004cj} on elliptic flow of charged particles in 16-23\% Au+Au collisions.}
\label{F2}
\end{figure}

Any dependence of the hydrodynamic evolution on integration step lengths
will be reflected on the freeze-out surface and consequently on 
observables like particle $p_T$ spectra, elliptic flow etc.  
In Fig.\ref{F2}, we have computed the elliptic flow for $\pi^-$   in  a b=6.5 fm Au+Au collision. The dashed and solid lines are $v_2$ in evolution with integration step lengths (i) $dx=dy=0.1fm,d\tau=0.01 fm$ and (ii) $dx=dy=0.2fm,d\tau=0.02 fm$ respectively. We have shown results both for ideal (black lines) and viscous fluid with two values of viscosity $\eta/s$=0.04 (the red lines) and 0.08 (the blue lines). 
The freeze-out temperature is assumed to be $T_F$=150 MeV. Both in ideal and in viscous dynamics, $v_2$ depend marginally on integration step lenghs.  Elliptic flow  
in evolution with  $dx=dy=0.1fm,d\tau=0.01 fm$ is $\sim$ 10\% more than in evolution with $dx=dy=0.2fm,d\tau=0.02 fm$. 
Elliptic flow is very sensitive observable. Stablity of $v_2$ 
imply that the hydrodynamic evolution, whether computed with integration step lengths  (i) $dx=dy=0.1fm,d\tau=0.01 fm$ or (ii) $dx=dy=0.2fm,d\tau=0.02 fm$,  
lead to nearly identical freeze-out surfaces. 

\begin{figure}[ht]
\includegraphics[bb=46 316 534 795
,width=0.7\linewidth,clip]{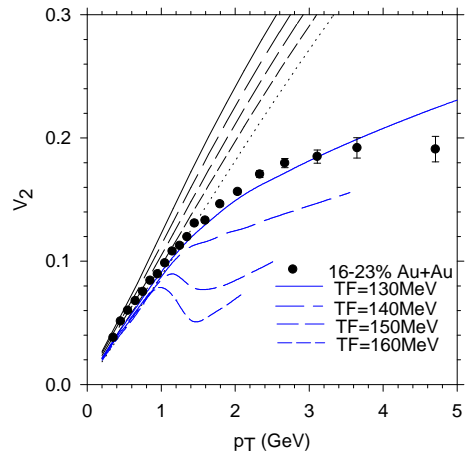} 
\caption{  $p_T$ dependence of elliptic flow in ideal (the black lines) and in minimal viscous (the blue lines) dynamics  for  $T_F$=130,140,150 and 160 MeV in a b=6.5 fm Au+Au collision. 
The filled circles are the PHENIX data \cite{Adler:2004cj} on the elliptic flow in 
16-23\% Au+Au collisions.}
\label{F3}
\end{figure}

\begin{figure}[ht]
\includegraphics[bb=31 291 524 770
,width=0.7\linewidth,clip]{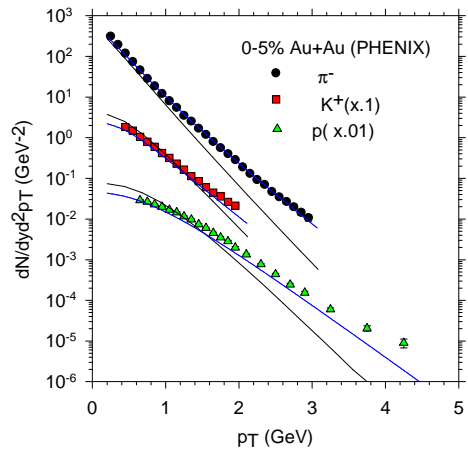}
\caption{PHENIX data \cite{Adler:2003cb} on the transverse momentum dependence $\pi^-$, $K^+$ and proton in 0-5\% Au+Au collisions are shown. The blue and black  lines are predictions from minimally viscous hydrodynamics and from ideal hydrodynamics in a b=2.3 fm Au+Au collision.}
\label{F4}
\end{figure}

Effect of viscosity on elliptic flow is also evident in Fig.\ref{F2}. Under identical conditions elliptic flow is less in viscous dynamics than in ideal dynamics. More viscous is the fluid less is the elliptic flow.
b=6.5 fm collision roughly corresponds to 13-26\% Au+Au collision. 
In Fig.\ref{F2}, PHENIX data \cite{Adler:2004cj} on the elliptic flow in 16-23\% Au+Au collisions are shown (the filled circles). As mentioned earlier, ideal fluid dynamics fails to explain the data beyond $p_T\approx$1.5 GeV. Data are not explained in 
minimally viscous dynamics also. ADS/CFT lower bound on viscosity $\eta/s$=0.08 reduce the elliptic flow more than necessary. Reduction is less for still lower viscosity and
PHENIX data are well explained up to $p_T\sim$3 GeV for viscosity $\eta/s$=0.04. A  similar result is also obtained in
\cite{Romatschke:2007mq}. Minimum bias STAR data seems
to favor smaller viscosity ($\eta/s$=0.03), lower than the ADS/CFT bound.

In a viscous fluid, elliptic flow depend strongly on the freeze-out
temperature.
For the minimally viscous fluid ($\eta/s$=0.08), PHENIX data could be explained if the hadronic freeze-out occur at temperature lower than 150 MeV. The shear stress tensors will evolve for longer duration to lower values and contribute less to the non-equilibrium distribution function and to the elliptic flow. 
In Fig.\ref{F3}, the elliptic flow for freeze-out temperatures  $T_F$=130,140,150 and 160 MeV are shown. Curves are drawn up to $p_T$ such that $\frac{dN^{neq}}{dyd^2p_T} < \frac{dN^{neq}}{dyd^2p_T}$. 
As expected,
elliptic flow increases as $T_F$ decreases. $p_T$ range over which viscous dynamics remain valid is also extended. For $T_F$=130 MeV, minimally viscous dynamics reproduce the
PHENIX data in 16-23\% Au+Au collisions, well  up to $p_T$=3.6 GeV.  However, viscous dynamics do not reproduce saturaton. At large $p_T$   jets are important. Without the effect of jets accounted for, possibly saturation of elliptic flow  could not be explained. For comparison, we have also shown the elliptic flow in ideal dynamics.  In ideal dynamics freeze-out dependence is compartively less.
We may mention that fit to elliptic flow data in other centrality ranges of collisions is not as good as shown in Fig.\ref{F3}. Minimally viscous dynamics produces less elliptic flow  in very central collisions (0-5\%, 5-10\%) collisions. In less central collisions data are reasonably well reproduced.

Viscous hydrodynamics with minimal viscosity $\eta/s$=0.08
also reproduces the transverse momentum distribution of identified particles.  In Fig.\ref{F4}, we have shown the PHENIX data \cite{Adler:2003cb} on 
transverse momentum distribution of $\pi^-$, $K^+$ and protons
in 0-5\% Au+Au collisions. 
The blue lines are viscous hydrodynamics predictions in a b=2.3 fm Au+Au collision. Freeze-outtemperature is $T_F$=130 MeV. Minimally viscous hydrodynamics ($\eta/s$=0.08) reproduces the $\pi^-$, $K^+$ and proton spectra    up to $p_T$=3 GeV. A comparable fit could not be obtained in ideal dynamics.

To summarise, in causal dissipative hydrodynamics, we have studied the effect of viscosity on elliptic flow. Under identical conditions, elliptic flow is less in viscous dynamics than in ideal dynamics. For minimally viscous fluid ($\eta/s$=0.08), PHENIX data on elliptic flow in 16-23\% Au+Au collisions are well explained up to $p_T\approx$3.6 GeV. This is to be contrasted with ideal dynamics which can explain the flow up to $p_T\approx$1.5 GeV only. Minimally viscous fluid also explains the $p_T$ spectra of $\pi^-$, $K^+$ and protons, much better than in ideal dynamics.


\end{document}